%% file: anoimp.tex
\documentclass[12pt]{article}

\title{}

\makeatletter
\long\def\@makemyfntext#1{$^{\rm *}\ $ #1}

\long\def\@myfootnotetext#1{\insert\footins{\footnotesize
    \interlinepenalty\interfootnotelinepenalty 
    \splittopskip\footnotesep
    \splitmaxdepth \dp\strutbox \floatingpenalty \@MM
    \hsize\columnwidth \@parboxrestore
   \edef\@currentlabel{\csname p@footnote\endcsname\@thefnmark}\@makemyfntext
    {\rule{\z@}{\footnotesep}\ignorespaces
      #1\strut}}}

\def\myfootnotetext{\@ifnextchar
     [{\@xfootnotenext}{\xdef\@thefnmark{\thempfn}\@myfootnotetext}}
\makeatother

\input macros

\newcommand{\allb}{\all^b}
\newcommand{\someb}{\some^b}

\newcommand{\av}{anonymous variables} %
\newcommand{\Av}{Anonymous variables} 

\begin{document}
	
\begin{center}
{\Large {\bf Anonymous Variables in Imperative Languages}}
\\[20pt] 
{Keehang Kwon\\
 Faculty of Computer Engineering, DongA  University\\
 khkwon@dau.ac.kr }
\end{center}
	
\noindent {\bf Abstract}: 

In this paper, we bring anonymous variables into imperative languages.
Anonymous variables represent don't-care values and have proven useful
in logic programming. To bring the same level of benefits into imperative languages, 
we describe an extension to C wth anonymous variables.




\section{Introduction}\label{sec:intro}

The notion of \av\   was introduced in logic programming. 
Anonymous variables represent don't-care values. As we shall see later, they
 provide some convenience to programming.
 This paper aims to bring \av\
 into imperative languages. Thus we allow the symbol $\_$ which denotes an anonymous variable.
 To see some use of \av, let us consider the following procedure
 which produces the amount of the tuition of a student $x$ with major $m$.

\begin{example}
   $\all x \all m$ tuition(x,m) = 		\\
	\>	case medical : amount = \$10K;\\
	\>	case english : amount = \$5K;\\
	\>	case physics : amount = \$5K;
\end{example}

Note that the above program is independent of $x$. To represent this, we replace the above with

\begin{example}
    $\all m$    tuition(\_,m) = 		\\
	\>	case medical : amount = \$10K;\\
	\>	case english : amount = \$5K;\\
	\>	case physics : amount = \$5K;
\end{example}
\noindent which is an abbreviation of 

\begin{example}
       $\allb x \all m$ tuition(x,m) = 		\\
	\>	case medical : amount = \$10K;\\
	\>	case english : amount = \$5K;\\
	\>	case physics : amount = \$5K;
\end{example}
\noindent where $\allb x$ is called a $blind$ universal quantifier\footnote{This concept
 was originally introduced in \cite{Jap03}, but with different notations.
For example, the  blind universal quantifier is denoted by $\all x$.}.
The main difference between $\allb x$  and $\all x$ is that, in the former, the instantiation of
$x$ will $not$ be visible to the user and will $not$ 
be recorded in the execution trace, while in the latter, it will.

 Now consider the following procedure call.

\[ tuition(kim,medical);print(amount). \]

\noindent 
Then the machine will print ``\$10K" as usual.

We also introduce its counterpart 
$\someb x_i\ p(x_1,\ldots,x_n)$ where $p(x_1,\ldots,x_n)$ is a
procedure call.  In this case $x_i$ becomes an anonymous variable.

Implementing \av\ is not too difficult.  
Below we describe a modest method to bring \av\ into imperative language. 
During execution \av\ will be replaced by some value. Choosing the proper value for
\av\   is often not trivial. Typically \av\ will be replaced by uninstantiated variables. These variables will be instanced later when
enough information is gathered. This process is typically known as $unification$.
Unification process will not be described here and we refer \cite{MNPS91} to the reader.

\section{The Language}\label{sec:logic}

The language is a subset of the core (untyped) C
 with some extensions. It is described
by $G$- and $D$-formulas given by the syntax rules below:
\begin{exmple}
\>$G ::=$ \>   $true \sep p(x_1,\ldots,x_n) \sep \some x p(x_1,\ldots,x_n) \sep x = E \sep  G;G  $ \\  
\>$D ::=$ \>  $ A = G\ \sep \all x\ D  \sep \allb x\ D \sep D \land D $\\
\end{exmple}
\noindent
 In the above, 
$A$  represents a head of an atomic procedure definition of the form $p(x_1,\ldots,x_n)$. 
A $D$-formula is a set
of procedure declarations. 

In the execution, a $G$-formula will function as a statement 
and a set of $D$-formulas  enhanced with the
machine state (a set of variable-value bindings) will constitute  a program.
Thus, a program is a union of two disjoint sets, \ie, $\{ D_1,\ldots,D_n \} \cup \theta$
where  $\theta$ represents the machine state.
$\theta$ is initially  empty  and will be updated dynamically during execution
via the assignment statements. 

 We will  present an interpreter for our language via natural semantics \cite{Khan87}.
It alternates between 
 the  execution phase 
and the backchaining phase.  
In  the  execution phase (denoted by $ex(\Pscr,G,\Pscr')$), it  
executes a statement $G$  with respect to
 $\Pscr$ and
produce a new program $\Pscr'$
by reducing $G$ 
to simpler forms. The rules
(7)-(10) deal with this phase. 
If $G$ becomes a procedure call, the machine switches to the backchaining mode. This is encoded in the rule (6). 
In the backchaining mode (denoted by $bc(D,\Pscr,A,\Pscr')$), the interpreter tries 
to find a matching procedure  for a procedure call $A$ inside the module $D$
 by decomposing $D$ into a smaller unit (via rule (4)-(5)) and
 reducing $D$ to  its instance
 (via rule (2),(3)) and then backchaining on the resulting 
definition (via rule (1)).
 To be specific, the rule (2) basically deals with argument passing: it eliminates the universal quantifier $x$ in $\all x D$
by picking a value $t$ for
$x$ so that the resulting instantiation,  $[t/x]D$, matches the procedure call $A$.
 The notation $S$\ seqand\ $R$ denotes the  sequential execution of two tasks. To be precise, it denotes
the following: execute $S$ and execute
$R$ sequentially. It is considered a success if both executions succeed.
Similarly, the notation $S$\ parand\ $R$ denotes the  parallel execution of two tasks. To be precise, it denotes
the following: execute $S$ and execute
$R$  in any order.  It is considered a success if both executions succeed.
The notation $S \leftarrow R$ denotes  reverse implication, \ie, $R \rightarrow S$.

\begin{defn}\label{def:semantics}
Let $G$ be a statement and let $\Pscr$ be a program.
Then the notion of   executing $\lb \Pscr,G \rb$ and producing a new
program $\Pscr'$-- $ex(\Pscr,G,\Pscr')$ --
 is defined as follows:

\begin{numberedlist}

\item    $bc((A = G_1),\Pscr,A,\Pscr_1)\ \leftarrow$  \\
 $ex(\Pscr,G_1,\Pscr_1)$. \% A matching procedure for $A$ is found.

\item    $bc(\all x D,\Pscr,A,\Pscr_1,)\ \leftarrow$  \\
  $bc([t/x]D,\Pscr, A,\Pscr_1)$. \% argument passing. Instantiation $(x,t)$ will  be recorded.

\item    $bc(\allb x D,\Pscr,A,\Pscr_1,)\ \leftarrow$  \\
  $bc([t/x]D,\Pscr, A,\Pscr_1)$. \% argument passing. Instantiation $(x,t)$ will not be recorded.

\item    $bc( D_1\land D_2,\Pscr,A,\Pscr_1)\ \leftarrow$  \\
  $bc(D_1,\Pscr, A,\Pscr_1)$. \% look for  a matching procedure in $D_1$.

\item    $bc( D_1\land D_2,\Pscr,A,\Pscr_1)\ \leftarrow$  \\
  $bc(D_2,\Pscr, A,\Pscr_1)$. \% look for a matching procedure in $D_2$

\item    $ex(\Pscr,p(x_1,\ldots,x_n),\Pscr_1)\ \leftarrow$    $(D \in \Pscr)$ parand $bc(D,\Pscr, A,\Pscr_1)$. \% $p(x_1,\ldots,x_n)$ is a procedure call

\item    $ex(\Pscr, \someb x_i p(x_1,\ldots,x_n),\Pscr_1)\ \leftarrow$   
$ex(\Pscr, [t/x_i]p(x_1,\ldots,x_n),\Pscr_1)$. \% $x_i$ is an anonymous variable.

\item  $ex(\Pscr,true,\Pscr)$. \% True is always a success.



\item  $ex(\Pscr,x = E,\Pscr\uplus \{ \lb x,E' \rb \}) \leftarrow$ $eval(\Pscr,E,E')$. \\
 \% In the assignment statement, it evaluates $E$ to get $E'$.
  The symbol $\uplus$ denotes a set union but $\lb x,V\rb$ in $\Pscr$ will be replaced by $\lb x,E' \rb$.

\item  $ex(\Pscr,G_1; G_2,\Pscr_2)\ \leftarrow$ \\
  $ex(\Pscr,G_1,\Pscr_1)$  seqand  $ex(\Pscr_1,G_2,\Pscr_2)$.
\%  a sequential composition

\end{numberedlist}
\end{defn}

\noindent
If $ex(\Pscr,G,\Pscr_1)$ has no derivation, then the interpreter returns  the failure.

\section{Examples }\label{sec:modules}

Let us consider again the example in the Introduction section.

\begin{example}
   $\all x \all m$ tuition(x,m) = 		\\
	\>	case medical : amount = \$10K;\\
	\>	case english : amount = \$5K;\\
	\>	case physics : amount = \$5K;
\end{example}

\noindent Now consider the following procedure call.

\[  tuition(\_,medical);print(amount). \]

Note that \_ is used in the above, as there is no need to specify a student. 
The above can be understood as an abbreviation of 

\[  \someb x\ tuition(x,medical);print(amount). \]

\section{Conclusion}\label{sec:conc}

In this paper, we have presented a notion of \av\  in the setting of
imperative languages.  We introduce $\allb$ for \av\ in procedure declarations and
$\someb$ for \av\ in procedure calls. \Av\ provide some convenience to programmers.

\bibliographystyle{ieicetr}



\end{document}

%% file: macros.tex
\newenvironment{numberedlist}
{\begin{list}{\makebox[20pt]{\hss(\arabic{itemno})\enspace}}
             {\usecounter{itemno}\labelwidth 20pt}}{\end{list}}

\newcounter{itemno}

\newcounter{itemno1}

\newcounter{itemno2}

\newcounter{exno}

\newcounter{defno}

\newenvironment{defn}{\refstepcounter{defno}\medskip \noindent {\bf
Definition \thedefno.\ }}{\medskip}

\newcommand{\sep}{\;\vert\;}

\newcommand{\oprove}{\vdash\kern-.6em\lower.7ex\hbox{$\scriptstyle O$}\,}

\newcommand{\Pscr}{{\cal P}}

\newcommand{\pderivation}{{\cal P}\kern -.1em\hbox{\rm -derivation}}
\newcommand{\pderivationl}{{\cal P}\kern -.1em\hbox{\em -derivation}}
\newcommand{\pderivable}{{\cal P}\kern -.1em\hbox{\rm -derivable}}
\newcommand{\pderivablel}{{\cal P}\kern -.1em\hbox{\em -derivable}}
\newcommand{\pderivations}{{\cal P}\kern -.1em\hbox{\rm -derivations}}
\newcommand{\pderivability}{{\cal P}\kern -.1em\hbox{\rm -derivability}}

\newcommand{\all}{\forall}
\newcommand{\some}{\exists}

\newcommand{\ie}{{\em i.e.}}

\newsavebox{\lpartfig}
\newsavebox{\rpartfig}

\newenvironment{exmple}{
 \begingroup \begin{tabbing} \hspace{2em}\= \hspace{3em}\= \hspace{3em}\=
\hspace{3em}\= \hspace{3em}\= \hspace{3em}\= \kill}{
 \end{tabbing}\endgroup}

\newenvironment{example}{
\begingroup  \begin{tabbing} \hspace{2em}\= \hspace{3em}\= \hspace{3em}\=
\hspace{3em}\= \hspace{3em}\= \hspace{3em}\= \hspace{3em}\= \hspace{3em}\= 
\hspace{3em}\= \hspace{3em}\= \hspace{3em}\= \hspace{3em}\= \kill}{
 \end{tabbing} \endgroup }

\newcommand{\lb}{\langle}
\newcommand{\rb}{\rangle}

     {\\* \hspace*{\fill} \end{trivlist}}